\documentclass[]{article}
\usepackage[utf8]{inputenc}
\usepackage{amsmath}
\usepackage{amsfonts}
\usepackage{amssymb}
\usepackage{hyperref}
\usepackage{graphicx}
\usepackage{caption}
\usepackage{subcaption}
\usepackage{float}
\usepackage{cite}
\usepackage{color}
\usepackage[left=2.2cm,right=2.2cm,top=2.5cm,bottom=2.5cm]{geometry}
\usepackage[affil-it]{authblk}

\title{Bohr-van Leeuwen theorem in non-commutative space}
\author{Shovon Biswas$^{a}$\thanks{shovon432@gmail.com}}
\affil{\small{Department of Electrical and Electronic Engineering, Bangladesh University of Engineering and Technology, Dhaka 1000, Bangladesh $^a$}}
\date{}

\begin{document}

\maketitle
\begin{abstract}
    Bohr-van Leeuwen theorem has been studied in non-commutative space where the space coordinates do not commute. It has been found that in non-commutative space Bohr-van Leeuwen theorem, in general, is not satisfied  and a classical treatment of the partition function of charged particles in a magnetic field gives rise to non zero magnetization.\newline
    
\end{abstract}

\section{Introduction}
Due to the appearance of non-commutativity in
string theories \cite{ref1} and the observation that ordinary and non-commutative gauge theories are equivalent  \cite{ref2} there have been much studies of physics on a non-commutative space recently. For example, quantum hall effect and Landau diamagnetism in non-commutative space have been discussed in references \cite{dyl2,dulat} and \cite{dyl1,gam2}. Hydrogen atom spectrum and Lamb shift is discussed in references \cite{kang,masud,masud2}. A discussion on non-commutative quantum mechanics and non-commutative classical mechanics can be found in references \cite{gamboa,gam3,djema,romero,class,therm}.\\\\
The non-commutative space can be characterized by the following commutation relations:
\begin{align}
    \begin{split}
        [\hat{X}_j,\hat{X}_k] &=i\hbar\theta_{jk}\\
        [X_j,P_k] &= i\hbar\delta_{jk} \\
        [P_j,P_k] &= 0, \quad j,k=1,2,3
    \end{split}
\end{align}
where $\{\theta_{ij}\}$ is the totally antisymmetric matrix which represents the non-commutative property of the coordinates on a non-commutative space. 
Non-commutative operators $(\hat{X}_j,\hat{P}_j)$ can be expressed in terms of ordinary coordinate and momentum operators $(X_j,P_j)$ as \cite{li,djema}:
\begin{align}
    \begin{split}
        \hat{X_j} &= X_j -\frac{1}{2} 
        \theta_{jk}P_k, \quad k=1,2,3.\\
        \hat{P_j} &=P_j
    \end{split}
\end{align}
\\\\
Here, $(X_j,P_j)$ satisfy the usual commutation relations:
\begin{align}
\begin{split}
    [X_j,X_k] &=0 \\
    [X_j,P_k] &= i\hbar\delta_{jk} \\
    [P_j,P_k] &= 0, \quad j,k=1,2,3
\end{split}
\end{align}
It should be noted that the representations of non-commutative coordinates and momenta in terms of ordinary ones given in (2) have non trivial impacts. For example, the Schrodinger equation in the non-commutative space takes the form of the ordinary one but with a shifted potential having highly non-trivial consequences\cite{gam3}.\\\\
Bohr-van Leeuwen theorem\cite{van}, on the other hand, is one of the very well known theorems of classical physics stating the nonexistence of magnetism in classical statistics. When statistical mechanics and classical mechanics are applied consistently, the thermal average of the magnetization is always zero\cite{aha}. This makes magnetism solely a quantum mechanical effect. In this work, based on the assumption that the phase space has a symplectic structure consistent with the rules of commutation of the non-commutative quantum mechanics given in (1), it will be proved that Bohr-van Leuween theorem is violated when classical partition function is written with respect to non-commutative phase space variables.\\\\
The organization of this paper is as follows. In section 2 the classical Bohr-van Leeuwen theorem is reviewed. In section 3 the effect of non-commutivity of space coordinates on Bohr-van Leeuwen theorem is discussed.  

\section{Review of Bohr-van Leeuwen Theorem }
As noted above the content of the Bohr–van Leeuwen theorem is the nonexistence of magnetism in classical statistics\cite{schwabl}. To prove the theorem let us evaluate the classical partition function for $N$ charged particles in the electromagnetic field \cite{schwabl}
\begin{equation}
    Z=\frac{\int d^{3N}p d^{3N}x}{(2\pi\hbar)^{3N}N!}e^{-\beta H\left( \{ \mathbf{p}^{(i)}-\frac{e}{c}\mathbf{A(\mathbf{x}^{(i)})}\},\{\mathbf{x}^{(i)}\}\right)}
\end{equation}
Here $ H\left( \{ \mathbf{p}^{(i)}-\frac{e}{c}\mathbf{A(\mathbf{x}^{(i)})}\},\{\mathbf{x}^{(i)}\}\right)$ is the Hamiltonian which can be written as
\begin{equation}
     H=\sum_{i=1}^{N}\frac{1}{2m}\left( \mathbf{p}^{(i)}-\frac{e}{c}\mathbf{A(\mathbf{x}^{(i)}} \right)^2+W_{coul}
\end{equation}
where $\mathbf{A}$ is the vector potential and  $^{(i)}$ stands for $i$th particle. The last term in equation (5) stands for the Coulomb interaction of the particles with each other. If we substitute
\begin{equation}
    \mathbf{p'}^{(i)}=\mathbf{p}^{(i)}-\frac{e}{c}\mathbf{A(\mathbf{x}^{(i)})}
\end{equation}
$Z$ becomes independent of $\mathbf{A}$ and thus of $\mathbf{B}$. With the free energy
\begin{equation}
     F=-\frac{1}{\beta}\ln Z
\end{equation} this leads to the magnetization
\begin{equation}
  \mathbf{M}=-\frac{\partial F}{\partial \mathbf{B}}=0.   
\end{equation}
Thus there can be no magnetism in classical physics.

\section{ Bohr-van Leeuwen Theorem in Non-commutative Space}
To study Bohr-van Leeuwen theorem in non-commutative space (NCS), we assume that the passage between NC classical mechanics and NC quantum mechanics can be realized via the following generalized Dirac quantization condition\cite{class,therm}
\begin{equation}
    \{f,g\}=\frac{1}{i\hbar}[F,G]
\end{equation}
Where $(F,G)$ stands for the operator associated with classical observables $(f,g)$ and $\{,\}$ stands for Poisson bracket. Using this rule we obtain from (1)
\begin{align}
    \begin{split}
        \{\hat{x}_j,\hat{x}_k\}&=\theta_{jk}\\
         \{\hat{x}_j,\hat{p}_k\}&=\delta_{jk}\\
          \{\hat{p}_j,\hat{p}_k\}&=0\\
    \end{split}
\end{align}
It is easy to see that the following representation of NC coordinates $(\hat{x}_j,\hat{p}_j)$ in terms of commutative coordinates $(x_j,p_j)$ satisfies relations given in (10)
\begin{align}
    \begin{split}
        \hat{x_j} &= x_j -\frac{1}{2} 
        \theta_{jk}p_k \\
        \hat{p_j} &=p_j, \quad j,k=1,2,3.
    \end{split}
\end{align}
Now following the proposal that non-commutative observable $F^{NC}$ corresponding to the observable $F(x_,p)$ in commutative space can be defined by \cite{kang,masud} 
\begin{equation}
     F^{NC}=F(\hat{x}_j,\hat{p}_j)
\end{equation}
we write the partition function (4) in terms of NC coordinates in non-commutative space as
\begin{equation}
    Z^{NC}=\frac{\int d^{3N}\hat{p} d^{3N}\hat{x}}{(2\pi\hbar)^{3N}N!}e^{-\beta H_{NC}\left( \{ \mathbf{\hat{p}}^{(i)}-\frac{e}{c}\mathbf{A(\mathbf{\hat{x}}^{(i)})}\},\{\mathbf{\hat{x}}^{(i)}\}\right)}
\end{equation}
Using the representation (11), partition function (13) in NCS can be expressed in terms of ordinary coordinates as follows
\begin{eqnarray}
    Z^{NC}=\frac{\int d^{3N}p d^{3N}x }{(2\pi\hbar)^{3N}N!}e^{-\beta H\left( \{ \mathbf{p}^{(i)}-\frac{e}{c}\mathbf{A}(\mathbf{x}^{(i)}-\frac{1}{2}\mathbf{\tilde{p}}^{(i)})\},\{\mathbf{x}^{(i)}-\frac{1}{2}\mathbf{\tilde{p}}^{(i)})\}\right)}
\end{eqnarray}
where $\mathbf{\tilde{p}}_j=\theta_{jk}p_k$. Now, since $\mathbf{A}$ depends on momenta in (13) one can not, in general, make a simple substitution like equation (6) to eliminate $\mathbf{A}$ from the partition function. Thus the partition function, in general, will be dependent on the magnetic field $\mathbf{B}$.\\\\
As an example let us calculate the partition function of $N$ non-interacting electrons in a uniform magnetic field on non-commutative space. Since the electrons are non-interacting the partition function on NCS for $N$ electrons can be written as 
\begin{equation}
    Z_N^{NC}=\frac{[Z_1^{NC}]^N}{N!}
\end{equation}
where $Z_1^{NC}$ is the single particle partition function in NCS given by
\begin{equation}
    Z_1^{NC}=\frac{1}{(2\pi\hbar)^3}\int e^{-\beta H_{NC}}d^3\hat{p} d^3 \hat{x}
\end{equation}
We let the magnetic field $\mathbf{B}$ be in the z-direction by
choosing the symmetric gauge $\mathbf{A}=B(-\frac{y}{2},\frac{x}{2},0)$. For a single electron the Hamiltonian (5) in ordinary phase space takes the form
\begin{equation}
     H=\frac{1}{2m}\left[\left(p_x-\frac{eB}{2c}y\right)^2+\left( p_y+\frac{eB}{2c}x \right) ^2\right]+\frac{p_z^2}{2m}
\end{equation}
Following the definition given in (12) and substituting the representations (11) in the Hamiltonian (17) we express the single particle partition function (16) in terms of ordinary coordinates and momenta as:
\begin{eqnarray}
     Z_1^{NC} &&=\frac{1}{(2\pi\hbar)^3}\int d^3p d^3 x e^{-\beta \left\{ \frac{1}{2m}\left[\left(\gamma p_x - \frac{eB}{2c}y \right)^2 +\left(\gamma p_y+\frac{eB}{2c}x \right)^2 \right] +\frac{p_z^2}{2m} \right\}}
\end{eqnarray}
where $\gamma=1-\frac{eB\theta}{4c}$ and $c$ is the velocity of light. We have assumed non-commutivity in $XY$ plane only. It should be noted that similar expressions have been found for the non-commutative Hamiltonian appearing inside the curly braces in equation (18) in references \cite{dyl1,dyl2} using star product definition. Performing the integration we obtain
\begin{equation}
     Z_1^{NC}=\frac{V}{\gamma^2\lambda}
\end{equation}
where $V$ is the volume and $\lambda=\frac{h}{\sqrt{2m\pi k_B T}}$ is the thermal de Broglie wavelength. Thus the $N$ particle partition function (14) becomes
\begin{equation}
     Z_N^{NC}=\frac{1}{N!}\left( \frac{V}{\gamma^2\lambda} \right)^N
\end{equation}
Now through $\gamma=1-\frac{eB\theta}{4c}$ the partition function depends on the magnetic field $B$. Also it should be noted that in the critical value of $B=\frac{4c}{e\theta}$ the partition function diverges. Thus the thermodynamic limit is applicable below this value. In fact, at the critical value of $B$ the first two quadratic momenta terms vanish in equation (18). This system is different from the system under consideration and should be studied separately \cite{dyl2}. Using formulae (7) and (8) we find the magnetization in non-commutative space to be
\begin{equation}
     M_{NC}=\frac{2Nk_B Te \theta}{4c-e\theta B}
\end{equation}
Thus a non zero magnetization appears from classical Hamiltonian in non-commutative space indicating the violation of Bohr-van Leeuwen theorem. In case we switch off $\theta$, we have $M=0$ reproducing the classical result. Finally, using the formula for susceptibility\cite{schwabl} $\chi=-\frac{1}{V}\frac{\partial ^2 F}{\partial B^2}$, we find the expression for magnetic susceptibility of electrons in non-commutative space
\begin{equation}
     \chi_{NC}=-\frac{1}{V}\left[\frac{2Nk_BTe^2\theta^2}{(4c-e\theta B)^2}\right]
\end{equation}
Thus electrons in a uniform magnetic field in NCS exhibit diamagnetism.
\section{Conclusion}
We have found that Bohr-van Leeuwen theorem is not satisfied in non-commutative space defined by symplectic structure (10). This needs special attention because classical treatment of simplest systems can give rise to magnetism in non-commutative space even without the introduction of a spin degree of freedom. Such a scenario is completely absent in ordinary classical mechanics. As an example, we have analyzed the simple system of non-interacting electrons in a uniform magnetic field in non-commutative space and found that a classical treatment of the problem in NCS gives rise to diamagnetism of electrons. Finally, we note that the discussion presented here can be easily be extended to non-commutative phase space where the momenta just like as coordinates do not commute.
\section{Acknowledgement}
The author would like to thank Professor Dr M. Arshad Momen and Professor Dr M. Chaichian for their  valuable comments on the manuscript. The author expresses his gratitude to Mir Mehedi Faruk for his help to present the manuscript. 

\end{document}